\documentclass[aps,prl,twocolumn,showpacs,superscriptaddress]{revtex4-1}
\usepackage{amsmath,amssymb,graphicx}
\usepackage{times}
\usepackage[table]{xcolor}
\usepackage[colorlinks,citecolor=red]{hyperref}

\newcommand\ket[1]{\left|\textstyle{#1}\right\rangle}

\newcommand\braket[1]{\left\langle\textstyle#1\right\rangle}

\newcommand\down{\downarrow}
\newcommand\up{\uparrow}
\newcommand\bfA{\mathbf{A}}
\newcommand\tot{\mathrm{tot}}

\begin{document}

\title{Recurrent delocalization and quasi-equilibration of photons in coupled circuit QED systems}

\author{Myung-Joong Hwang}
\affiliation{Korea Insitute for Advanced Study, Seoul 02455, Korea}
\affiliation{Institut f\"{u}r Theoretische Physik, Albert-Einstein Allee 11, Universit\"{a}t Ulm, 89069 Ulm, Germany}

\author{M. S. Kim}
\affiliation{Korea Insitute for Advanced Study, Seoul 02455, Korea}
\affiliation{QOLS, Blackett Laboratory, Imperial College London, London SW7 2AZ, United Kingdom}

\author{Mahn-Soo Choi}
\email{choims@korea.ac.kr}
\affiliation{Department of Physics, Korea University, Seoul 02841, Korea}

\begin{abstract}
We explore the photon population dynamics in two coupled circuit QED systems. For a sufficiently weak inter-cavity photon hopping, as the photon-cavity coupling increases, the dynamics undergoes double transitions first from a delocalized to a localized phase and then from the localized to another delocalized phase.
The latter delocalized phase is distinguished from the former one; instead of oscillating between the two cavities, the photons rapidly \emph{quasi}-equilibrate over the two cavities.
These intrigues are attributed to an interplay between two qualitatively distinctive nonlinear behaviors of the circuit QED  systems in the utrastrong coupling regime, whose distinction has been widely overlooked.
\end{abstract}


\maketitle

A single quantum emitter strongly coupled to a quantized electromagnetic field can induce a significant interaction among photons~\cite{Imamoglu:1997gp}, which are usually very weakly interacting. The Jaynes-Cummings (JC) model~\cite{Jaynes:1963fa}, typically realized in a cavity QED system,  provides an intuitive understanding. The nonlinearity of the energy spectrum of the JC model causes an additional energy cost to put an extra photon into the cavity~\cite{Birnbaum:2005cj}, which gives rise to the effective interaction energy of photons.
When photons are allowed to hop between nearby cavities forming a so-called JC-Hubbard lattice, a new physics of strongly correlated photons emerges~\cite{Greentree:2006jg, Angelakis:2007ho, Hartmann:2006kv, Gerace:2009dp, Carusotto:2013gh}. In equilibrium, the JC-Hubbard lattice shows critical behaviors that resemble the physics of Bose-Hubbard model~\cite{Koch:2009hh, Schmidt:2009cs}. 

An interesting consequence of the competition between the qubit-cavity coupling and the photon hopping is a self-trapping transition~\cite{Schmidt:2010kr}, as recently observed in an experiment using two capacitively coupled superconducting transmission lines and transmon qubits~\cite{Raftery:2014jk}. The self-trapping transition in tunnel-coupled quantum systems occurs when an on-site interaction energy becomes so dominant that it prevents quantum tunneling through the tunnel barrier~\cite{Smerzi:1997fm,Albiez:2005ku,Abbarchi:2013it}. Likewise, when the nonlinearity induced by the qubit-cavity coupling exceeds the inter-cavity photon hopping, the photon population dynamics undergoes a sharp transition from a delocalized (tunneling) to a localized (self-trapping) regime~\cite{Schmidt:2010kr,Raftery:2014jk}.

Meanwhile, the qubit-cavity coupling that is comparable to a qubit transition frequency or a cavity frequency, the so-called ultrastrong coupling (USC), has been recently achieved in experiments~\cite{Ciuti05a,Gunter:2009gc, Niemczyk:2010gv, FornDiaz:2010by,Zazunov03a,Janvier15a}. In the USC regime, the rotating wave approximation leading to the JC model is not applicable, thus the total excitation number is not conserved~\cite{Braak:2011hc, Ashhab:2010eh,Hwang:2010jn, Casanova:2010kd}.
The counter-rotating (CR) terms, which are neglected in the rotating wave approximation, play a crucial role in the physics of strongly correlated photons induced by a light-matter interaction. Equilibrium studies on the JC-Hubbard lattice have shown that the USC leads to an Ising-type quantum phase transition and an exotic phase of light~\cite{Schiro:2012hr, Hwang:2013kk, Zhu:2013uc,Hartmann08a}. 

In this Letter, we explore the dynamics of strongly correlated photons in two coupled circuit QED systems in the USC regime.
We examine the phase diagram in the parameter space consisting of the qubit-cavity coupling and the inter-cavity photon hopping.
We find that as the photon-cavity coupling increases, the dynamics undergoes double transitions first from a delocalized to a localized phase and then from the localized to another delocalized phase.
Moreover, the latter phase is characterized by the quasi-equilibration of photon population, despite that the system is finite and closed.
We explain the results based on a competition between two qualitatively
distinctive nonlinear behaviors of the circuit QED systems in the USC regime:
One nonlinear regime, which is commonly associated with the
photon-blockade effect and responsible for the first
delocalization-localization transition, has been explored in various contexts in
previous works~\cite{Imamoglu:1997gp,Birnbaum:2005cj,Ridolfo:2012dt}. However, the other nonlinear regime, responsible for the second localization-delocalization transition and
the quasi-equilibration dynamics of photon population, has been widely
overlooked so far.
Interestingly, the same picture also explains the absence of the photon blockade in single photon transfer dynamics studied previously in Ref.~\cite{Felicetti:2014fu}.
We note that our findings can be observed by combining existing circuit QED technologies used for the JC dimer experiment~\cite{Raftery:2014jk} and for the realization of USC regime~\cite{Niemczyk:2010gv, Bourassa:2009gy, Nataf:2011ff}.

\begin{figure}
\begin{center}
\includegraphics[width=8cm]{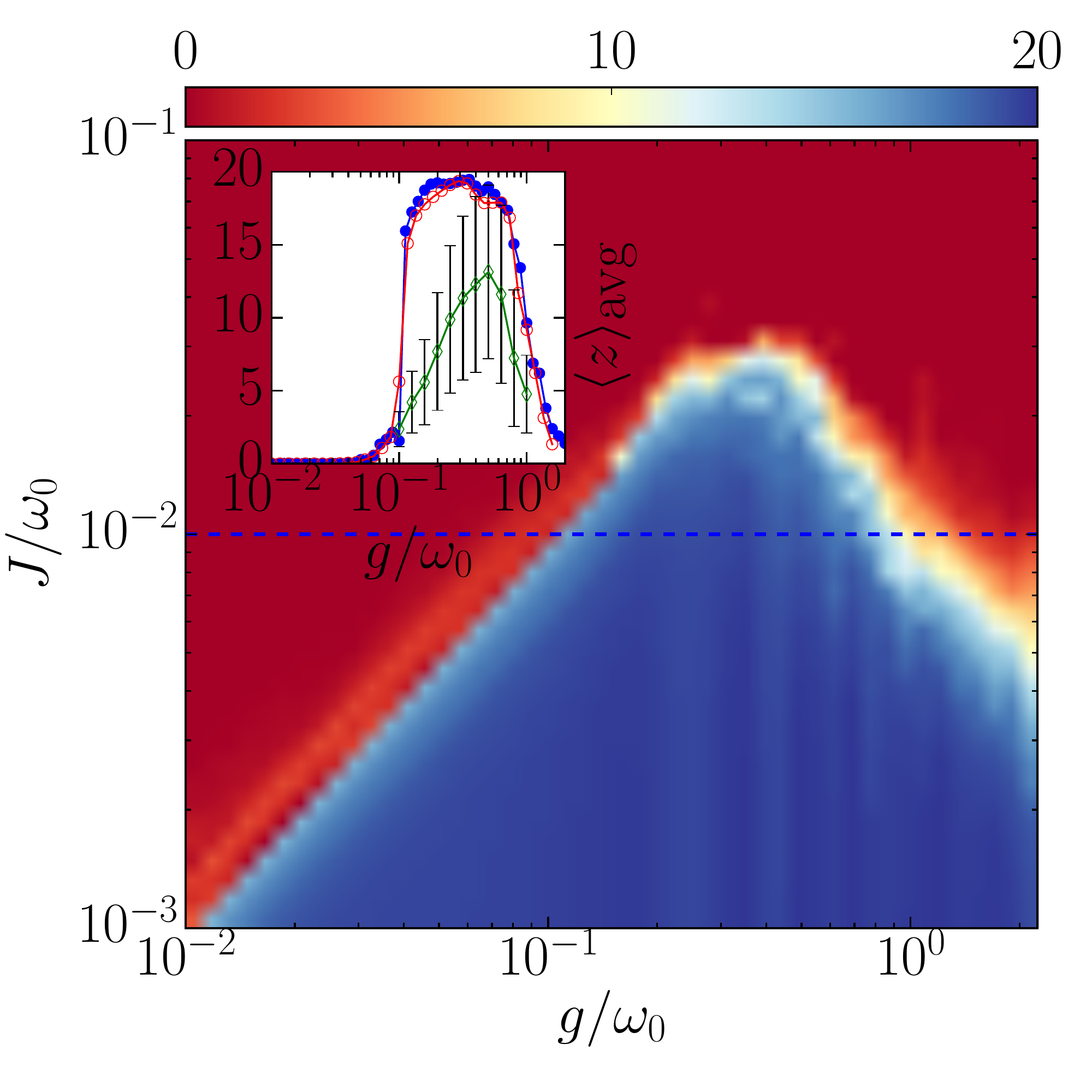}
\caption{\label{fig:1}Phase diagram for the average photon population imbalance $z_\mathrm{avg}$ for the initial state $\ket{20,\down}_L\ket{0,\down}_R$.
  (inset) $z_\mathrm{avg}$ at $J/\omega_0=0.01$ for different initial states and damping conditions: The filled circle is for the Fock state $\ket{20,\down}_L\ket{0,\down}_R$, the empty circle for the coherent state $\ket{\sqrt{20},\down}_L\ket{0,\down}_R$, and the empty diamond for $\ket{20,\down}_L\ket{0,\down}_R$ with finite cavity damping time $\tau_\gamma=10^4/\omega_0$ (averaged over 300 quantum trajectories).
}
\end{center}
\end{figure}

\paragraph{Model---}
Our system is described by the Rabi-dimer model,
\begin{equation}
\label{eq:1}
\hat H=\hat H^\mathrm{Rabi}_L + \hat H^\mathrm{Rabi}_R
-J (\hat a^\dagger_L \hat a_R+\hat a^\dagger_R \hat a_L)
\end{equation}
consisting of
the two Rabi interaction systems
\begin{equation}
\label{eq:2}
\hat H^\mathrm{Rabi}_{j=L,R} =\omega_0\hat a^\dagger_j \hat a_j+\frac{\Omega}{2}\hat \sigma_j^z-g(\hat a_j+\hat a_j^\dagger)\hat \sigma_j^x
\end{equation}
coupled to each other via the photon tunneling with amplitude $J$.
The left (L) and right (R) Rabi interaction systems are assumed to be identical with the cavity frequency $\omega_0$, the qubit transition frequency $\Omega$, and the qubit-cavity coupling strength $g$.
The coupling strength $J$ is assumed to be sufficiently weak ($J\ll\omega_0$) as in common experiments~\cite{Raftery:2014jk}. The operator $\hat a_j$  describes the field mode of the cavity $j$, and Pauli operators $\hat \sigma^{x,y,z}_j$ the qubits. Note that the Rabi Hamiltonian~(\ref{eq:2}) contains the CR terms,
$\hat a_i\hat \sigma^-_i+\hat a^\dagger_i\hat \sigma^+_i$
with
\begin{math}
\hat \sigma^{\pm}_i=(\hat \sigma_i^x \pm i\hat \sigma^y_i)/2,
\end{math}
in addition to the JC Hamiltonian. We focus our analysis on the resonant case, $\omega_0=\Omega$, where the effective photon-photon interaction is strongest for given coupling strengths. 

To investigate the photon-localization dynamics,
we suppose that the photons are initially localized in one cavity~\cite{Smerzi:1997fm,Albiez:2005ku,Abbarchi:2013it,Schmidt:2010kr,Raftery:2014jk}. Specifically, we mainly focus on the case where the initial state is of the particular type $\ket{\Psi^\tot(t=0)}=\ket{n_i,\down}_L\ket{0,\down}_R$ with $n_i>10$, where $|n,\sigma\rangle_j$ denotes a product state of $n$-photon Fock state and $\sigma=\up,\down$ qubit state in the cavity $j=L,R$.
A few remarks are in order: (i) An initial Fock state is just to simplify the
discussion. We have investigated the case of initial coherent states
\cite{Rempe87a,Gea-Banacloche90a} and the results are essentially the same
(see the inset of Fig.~\ref{fig:1}) within our parameter regime.
(ii) A relatively large number of initial photons ($n_i>10$ in our simulation)
is required because otherwise the localization time is known to become too
short (the localized phase disappears)~\cite{Schmidt:2010kr}.

We describe the photon localization-delocalization transition in terms of the \emph{unnormalized} photon population imbalance parameter, $z(t)=\langle\hat N_L(t)-\hat N_R(t)\rangle$ with $\hat N_j= \hat a_j^\dagger \hat a_j$, and its time-averaged value $z_\mathrm{avg}=\frac{1}{T}\int^T_0 z(t)$ with the ``observation'' time $T$.
Note that for $g/\omega_0\gtrsim1$ the total number of photons
\begin{math}
N_\text{tot}(t)=\langle\hat N_{L}(t)+\hat N_{R}(t)\rangle
\end{math}
can be significantly different from the initial number $n_i$ of photons,
because the CR terms
can generate (or destroy) a considerable number of photons from (or to) vacuum.
In such a regime
the \emph{normalized} photon population imbalance, $z_\mathrm{norm}(t)=\frac{N_L(t)-N_R(t)}{N_L(t)+N_R(t)}$, that has been commonly used to distinguish the localized and delocalized regime in the previous studies~\cite{Smerzi:1997fm,Albiez:2005ku,Abbarchi:2013it, Schmidt:2010kr, Raftery:2014jk}, can severely underestimate the imbalance.
In our simulation, we set $T=2\times 10^4/\omega_0$, which is sufficiently long in the parameter regime of interest (to be discussed in greater detail below) to distinguish the localized and delocalized phases.

\begin{figure}
\begin{center}
\includegraphics[width=\linewidth]{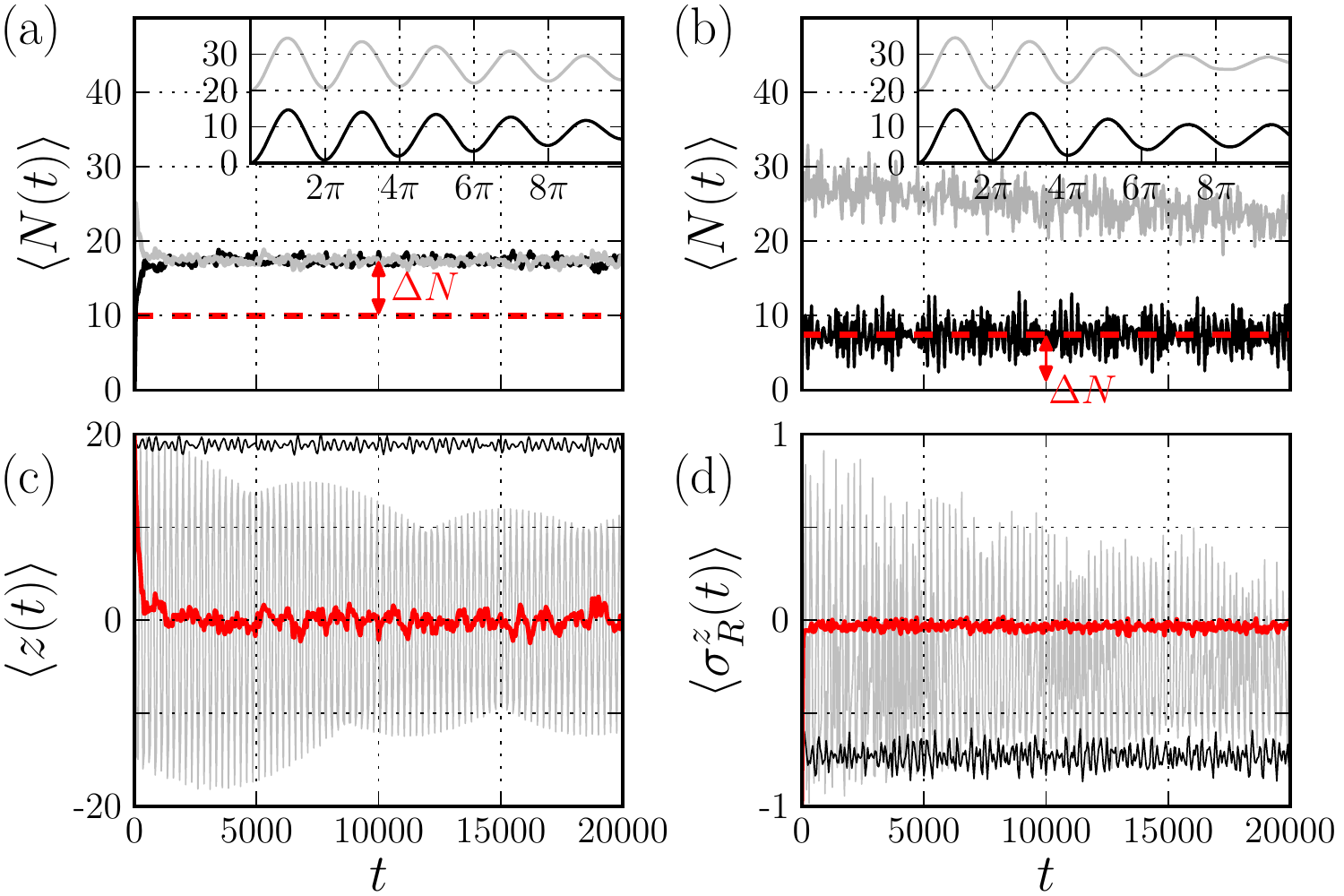}
\caption{\label{fig:2}(a) The photon population dynamics of a Rabi dimer with $J/\omega_0=0.01$ and $g/\omega_0=2$, where the gray (black) line corresponds to $\braket{N_{L(R)}(t)}$. (b) The photon population dynamics of a \emph{single} Rabi model with $g/\omega_0=2$, where the Gray (black) line corresponds to the initial state $\ket{n_i=20,\down}$ ($\ket{n_i=0,\down}$). (insets of (a) and (b)) The corresponding transient dynamics, with the number of vacuum-generated photons $\Delta N\sim7.5$. (c) The photon population imbalance $\braket{z(t)}$ and (d) the qubit polarization $\braket{\sigma_R^z(t)}$ in the right cavity for $J/\omega_0=0.01$ and $g/\omega_0=0.01~(\mathrm{Gray}),~0.2~(\mathrm{thin black}),~\mathrm{and}~2~(\mathrm{thick red})$.}
\end{center}
\end{figure}

\paragraph{Results---} Figure~\ref{fig:1} shows the phase diagram of the photon population dynamics in the $g$-$J$ space determined by $z_\text{avg}$.
Note that  for the Rabi-dimer both $g/\omega_0$ and $J/\omega_0$ become relevant even in the resonant case $\omega_{L(R)}=\Omega_{L(R)}$ whereas
for the JC-dimer at resonance the dynamics is solely governed by the ratio $J/g$.
The phase diagram exhibits a sharp distinction between the \emph{localized} regime, where the photons are self-trapped, and the \emph{delocalized} regime, where photons tunnel between two cavities.
More importantly, the phase boundary is nonmonotonic:
(i) There exists a critical value $J_c\approx 0.03\omega_0$ of $J$, above which photons are delocalized for all $g$.
Note that $J_c\approx0.03\omega_0$ is small enough for the tunneling Hamiltonian~(\ref{eq:2}) to be valid for the cavity-cavity coupling.
(ii) For $J<J_c$, as $g$ increases, the system undergoes recurrent transitions, first from a delocalized to localized at $g_{c1}$ and then from the delocalized to another localized phase at $g_{c2}$.
The critical value of the first transition scales as $g_{c1}\sim J\sqrt{n_i}$, as already shown in Ref.~\cite{Schmidt:2010kr}.
The second transition, on the other hand, happens at the critical value,  $g_{c2}\sim J_c/J$ (note that Fig.~\ref{fig:1} is in a logarithmic scale), which hardly depends on the initial photon number.

A semiclassical approach has been proposed in Ref.\cite{Schmidt:2010kr} that describes the first transition well. However, it breaks down for larger $g$ and completely misses the recurrent delocalization transition.
In passing, Ref.~\cite{Raftery:2014jk} ascribes the first delocalization-localization transition to a classical-to-quantum transition in the sense that the collapse-and-revival
emerges in the localized regime when either cavity is initially populated with a \emph{coherent state},
while the delocalized regime is characterized by classical oscillations.

 
Even more interesting is the dynamical characteristics of the second delocalized phase, clearly distinguished from the first one. To see this, let us turn to the photon number in each cavity $N_{L(R)}(t)$ and the photon population imbalance $z(t)$ in the USC regime, presented in Fig.~\ref{fig:2} (a) and (c).
The two cavities share almost the same number of photons at any time $t$ after the transient time dynamics; see the inset of Fig.~\ref{fig:2} (a). That is, the fluctuations around $z_\mathrm{avg}=0$ are highly suppressed. It is intriguing to find that the system starting from an imbalanced photon population distribution between the cavities rapidly equilibrates to share an equal number of photons, given that our model includes only two lattice sites without any contact to a bath.  The quasi-equilibration of photon dynamics is accompanied by the depolarization  of the qubits, $\braket{\sigma_{L(R)}^z(t)}\sim0$; see Fig.~\ref{fig:2} (d).
Experimentally, the qubit depolarization is a useful signature of the quasi-equilibration as the qubit state is usually easier to probe than the cavity photon number.

The effect is referred to as \emph{quasi}-equilibration
because a closed quantum system requires a recurrence of dynamics at a finite time $\tau_r$ ~\cite{Bocchieri:1957bn}.
However, it is remarkable that according to our numerical simulation $\tau_r\gg T=10^4\omega_0^{-1}$.
For the circuit QED system, the transmission line resonator and the transmon qubit have a few gigahertz frequency, while relaxation rates are typically in a few megahertz. Therefore, $\tau_r$ is much longer than any time scales relevant to the experiment, practically equivalent to $\tau_r\rightarrow\infty$. We also note an interesting resemblance to the recently predicted quasi-equilibration between two identical \textit{finite} quantum systems~\cite{Ponomarev:2011bi}, where a common temperature and small fluctuations around time average of any observables are key features. Our finding also provides a concrete example to the recent discussion of the equilibration in a closed quantum system~\cite{Ponomarev:2011bi,Reimann:2008hq,*Short:2011fq,*Short:2012dta}.

\paragraph{Discussions---}
We provide qualitative explanations for the dynamical features of the Rabi-dimer model based on the peculiar properties of the \emph{dressed} states of \emph{single} Rabi models.

At $g=0$, the eigenstates of the Rabi Hamiltonian $\hat H^\text{Rabi}$ [see Eq.~(\ref{eq:2})] is a product state $\ket{n}\ket{\up(\down)}$ of a cavity field mode and a qubit.
Namely, the field mode is decoupled from the qubit and its dynamical behavior is harmonic (i.e., linear).
As $g/\omega_0$ grows from zero until $g/\omega_0\lesssim 1$, the JC terms start to take effect and play a dominant role over the CR terms.
It leads to the coherent superposition of $\ket{n}\ket{\up}$ and $\ket{n+1}\ket{\down}$ and the corresponding $\sqrt{n}$-dependent splitting of eigenvalues. The nonlinearity in this range of coupling $g$ is thus characterized by the $\sqrt{n}$-dependence of the energy levels. The so-called JC nonlinearity induces an effective photon-photon interaction and is known to cause the photon-blockade (PB) effect~\cite{Schmidt:2010kr}.

In the opposite limit ($g/\omega_0\to\infty$ at resonance and $g/\sqrt{\omega_0\Omega}\to\infty$ in general), the $\frac{\Omega}{2}\hat\sigma_z$-term is negligible, and the eigenstates of $\hat H^\text{Rabi}$ have the form
\begin{math}
\ket{n,\pm g/\omega_0}\ket{\pm},
\end{math}
where $\ket{n,\alpha}=e^{\alpha \hat a^\dag - \alpha^*\hat a}\ket{n}$ for a complex number $\alpha$ is a displaced Fock state and $\ket{\pm}$ is a $\hat\sigma^x$-eigenstate. The field mode is thus linear again and the energy spectrum is harmonic~\cite{Feranchuk96a,Irish07a}.
One important difference (compared with the $g=0$ case) is that the photon number fluctuations in each eigenstate are huge (eventually diverge with $g$).
As $g/\omega_0$ decreases from the infinity to $g/\omega_0\gtrsim1$, the $\frac{\Omega}{2}\hat\sigma^z$-term tends to induce transitions between $\ket{n,\pm g/\omega_0}\ket{\pm}$, whose transition amplitude is determined by $\braket{n,-g/\omega_0|n,g/\omega_0}\propto e^{-2g^2/\omega_0^2}L_n(4g^2/\omega_0^2)$ where $L_n$ is the $n$-th Laguerre polynomial. The exponential suppression of the transition amplitude between states, $\ket{\pm}$, due to a state-dependent displacement of an oscillator, $\ket{\pm g/\omega_0}$, known as the Franck-Condon (FC) effect~\cite{Franck26a,Condon26a,Koch06a,ChoiMS07e}, leads to an exponential suppression of the energy splitting between $\ket{n,\pm g/\omega_0}\ket{\pm}$, and it governs the nonlinearity of the field mode in this range of $g$. It is stressed that in this range the CR terms play a crucial role and enable the vacuum to ``erupt'' a large number of photons.

The key observation in the above discussions is that one can expect two qualitatively distinctive nonlinear phases of the Rabi model and a transition between them at $g\sim\omega_0$ (the actual transition point may vary depending on details, such as $J$). We will refer to them as the photon-blockade (PB) and photon-eruption (PE) phases, respectively.
To examine the distinction more closely, we introduce
the photon number variance
\begin{math}
\chi = \langle \hat N^2\rangle-\langle \hat N\rangle^2
\end{math}
of the eigenstates and the level spacing variance
\begin{math}
\zeta = \frac{1}{K}\sum_kE_{k+1,k}^2
- \left(\frac{1}{K}\sum_kE_{k+1,k}\right)^2,
\end{math}
where $E_{kk'}\equiv E_k-E_{k'}$ are the differences between eigenenergies $E_k$ and $E_{k'}$ and $K$ are the number of considered eigenstates (ideally $K=\infty$). Naturally, $\zeta$ characterizes the \emph{nonlinearity} of the system.
One can distinguish the two distinct nonlinear phases as a competition between the nonlinearity $\zeta$ and the photon number fluctuation $\chi$. For $g/\omega\geq1$, the photon number variance of the eigenstates follows that of the coherent Fock state with a coherent amplitude of $g/\omega_0$; that is, $\chi\propto(g/\omega_0)^2$~\cite{deOliveira:1990gra}. As expected from the above discussions and illustrated in Fig.~3 (b), $\chi$ monotonically increases,  whereas $\zeta$ first increases but then decreases. For relatively small $g/\omega_0$, the nonlinearity dominates over the photon number fluctuations. It leads to the effective photon-photon interaction---hence the PB phase. For larger $g/\omega_0$, on the other hand,
the photon eruption due to the CR terms give rise to the large photon number fluctuations, which diminish the effective photon-photon interaction---hence the PE phase.

Now it is fairly straightforward to understand the phase diagram in Fig.~\ref{fig:1} and the corresponding photon population imbalance dynamics in Fig.~\ref{fig:2} of the Rabi dimer model.
For example, consider the case of $J/\omega_0=0.01$. In the weak-coupling regime ($g/\omega_0\ll 1$), the system is sufficiently linear and photons oscillate back and forth between the two cavities. As $g/\omega_0$ increases, the JC nonlinearity sets in and when $g/\omega_0>0.1$ photons are localized in one cavity due to the PB effect. As the coupling increases further so that $g/\omega_0>1$, the FC nonlinearity dominates and the system enters the PE phase. In this regime the photons are delocalized again.
Unlike the $g/\omega_0\ll1$ limit, however, the number of photons in each cavity does not oscillate in time but quasi-equilibrates over the two cavities.
We remark that the true $g/\omega_0\gg 1$ limit has not been observed in our actual simulation of the Rabi dimer because of the immense computational cost for large $g$. In such a limit the system becomes completely linear and should exhibit oscillations at a finite frequency (comparable to $J$).

\begin{figure}
\centering
\includegraphics[width=0.49\linewidth]{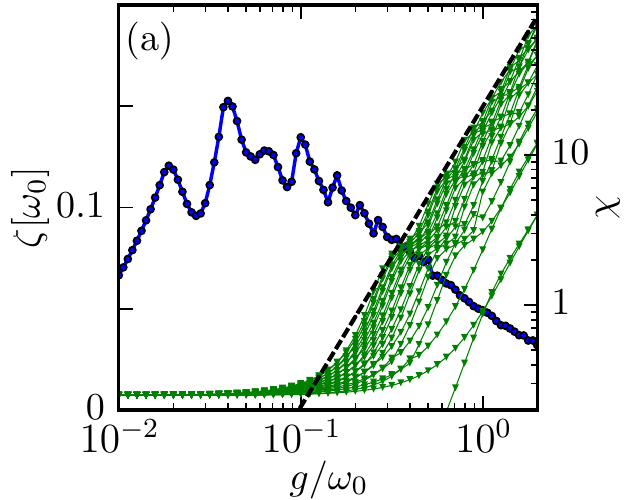}
\includegraphics[width=0.49\linewidth]{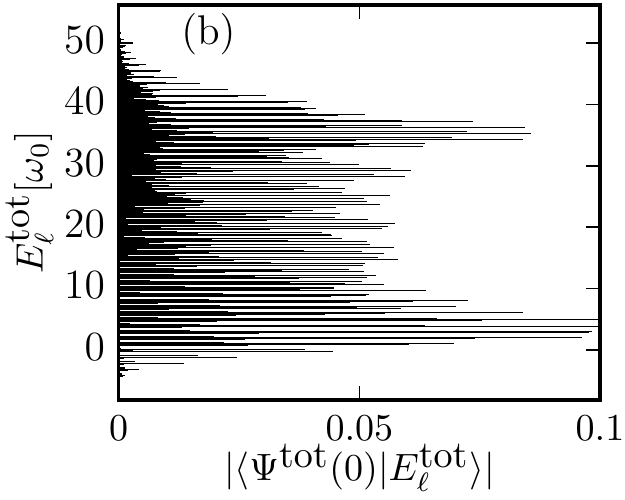}
\caption{\label{fig:3}(a) The level-spacing variance $\zeta$ (black filled circles, using the left vertical axis) for the 400 lowest levels and the photon number variance $\chi$ (solid lines, using the right vertical axis) for the $20$ lowest levels of the Rabi model. For $g/\omega_0>1$, the photon variance increases with $(g/\omega_0)^2$, as indicated by the dashed line. (b) The overlaps between an initial state $|\Psi(t=0)\rangle=|n_i,\down\rangle_L|0,\down\rangle_R$ and eigenstates of Rabi dimer Hamiltonian.}
\end{figure}

One remaining question is how the relatively simple system of a Rabi dimer can have a quasi-equilibration state.
To address the issue, we note that the initial localized Fock state $\ket{\Psi^\tot(0)}$ involves a wide range of eigenstates $\ket{E_\ell^\tot}$ (labeled by an integer index $\ell$) as demonstrated in Fig.~\ref{fig:3}~(b). Combined with the nonlinearity which makes the energy spectrum $E_\ell^\tot$ highly irregular, it leads to the unusually long recurrence time. Indeed, $\tau_r\gg T$ in both nonlinear phases (the PB and PE phases); see Fig.~\ref{fig:2}.

The crucial difference between the two nonlinear phases is the photon number fluctuations: In the PB phase the photon-hopping $J$ becomes irrelevant
while in the PE phase the enhanced photon number fluctuations enable $J$ to equilibrate photons over the two cavities.
To see this, we examine $\langle\hat N_j(t)\rangle$:
\begin{multline}
\label{eq:8}
\braket{\hat N_{j}(t)}=\sum_{\ell\ell'} e^{-iE_{\ell\ell'}t}
\braket{\Psi^\tot(0)|E^\tot_\ell}\braket{E^\tot_{\ell'}|\Psi^\tot(0)}
\\{}\times
\braket{E^\tot_\ell|\hat N_{j}|E^\tot_{\ell'}}.
\end{multline}
Recall that the spectrum involved in the sum is macroscopically wide and highly irregular. Then the off-diagonal terms with the fast oscillating factor $e^{-iE_{\ell\ell'}t}$ tend to cancel each other. At long time scales ($\omega_0^{-1}\ll t\ll\tau_r$), one thus expects
\begin{equation}
\label{eq:7}
\braket{\hat N_j(t)} \approx
\sum_{\ell} |\braket{\Psi^\tot(0)|\psi_\ell}|^2\braket{\psi_\ell|\hat N_{j}|\psi_{\ell}} = \text{const.}
\end{equation}
Since the system is symmetry under $L\leftrightarrow R$, the eigenstates $\ket{E^\tot_\ell}$ are either symmetric or antisymmetric. Therefore, one has
\begin{math}
\langle E^\tot_\ell| \hat N_{L}|E^\tot_\ell\rangle
=\langle E^\tot_\ell|\hat N_{R}|E^\tot_\ell\rangle
\end{math}
and hence
\begin{math}
\langle\hat N_L(t)\rangle =
\langle\hat N_R(t)\rangle
\end{math}
(i.e., quasi-equilibration) in the PE phase.
In the PB phase, the eigenstates $\ket{E^\tot_\ell}$ are highly localized on either cavity.
The overlap with the initial states $\braket{\Psi^\tot(0)|E^\tot_\ell}$ is negligible for those states $\ket{E^\tot_\ell}$ localized on the right cavity. It implies that $\langle N_L(t)\rangle \sim n_i$ and $\langle N_R(t)\rangle \sim 0$ in the PB phase.

So far we have ignored the $\bfA^2$-term due to the electromagnetic vector potential $\bfA$ of the cavity field \cite{Nataf10b,Tufarelli15a},
\begin{equation}
\hat H_A = D\sum_j(\hat a_j+\hat a_j^\dag)^2,
\end{equation}
where $D\propto g^2$.
Here we show that in circuit QED systems the term does not affect our results.
Note that the overall Hamiltonian $\hat{H}+\hat{H}_A$ is equivalent to the model in Eq.~(\ref{eq:1}) up to the unitary transformation
\begin{math}
\hat S
= \exp[r\sum_j\hat a_j^\dag \hat a_j^\dag - h.c.]
\end{math}
with $r$ defined by $e^{4r}=1+4D/\omega_0$.
But the parameters are renormalized as
\begin{math}
\omega_0 \to \tilde\omega_0=\omega_0e^{2r}
\end{math}
\begin{math}
g \to \tilde{g}=g e^{r}
\end{math} and
\begin{math}
J \to \tilde{J}=Je^{2r}.
\end{math}
Therefore the $\bfA^2$-term tends to decrease the reduced qubit-cavity coupling $\tilde{g}/\tilde\omega_0$ by factor $e^{-r}$ keeping the reduced cavity-cavity coupling $\tilde{J}/\tilde\omega_0$ the same. For the true atom-light coupling, the Thomas-Reiche-Kuhn (TRK) sum rule leads to $D/\omega_0>(g/\omega_0)^2$, and the dynamical features we have discussed above are very difficult to observe experimentally.
However, in circuit QED systems the underlying physics of qubit-cavity coupling is different, either the TRK sum rule does not apply or
the coupling $D$ has an additional suppression factor
\cite{Nataf10b}.
Therefore, the $\bfA^2$-term does not affect the relevant parameter region.


\paragraph{Remarks---} 
We have mainly focused on the unitary dynamics of the double Rabi systems.
In realistic experiments,
it is expected that the cavity damping time $\tau_\kappa$ (devided by $\langle{\hat N}\rangle$) or the qubit decoherence time $\tau_\phi$ sets the observation time $T$ for the recurrent delocalization and quasi-equilibration transition. Note that the recent state-of-the-art experiments \cite{Raftery:2014jk, Gunter:2009gc, Niemczyk:2010gv, FornDiaz:2010by, Bourassa:2009gy} have realized $\tau_\kappa,\tau_\phi>10^4/\omega_0$.
We have briefly examined the cavity damping effect based on the quantum jump approach \cite{Plenio98a} and taking into account the nontrivial interplay between the damping $1/\tau_\gamma$ and the strong coupling $g$.
As illustrated in the inset of Fig.~\ref{fig:1},
the localized phase becomes less prominent (as noted in Ref.\cite{Schmidt:2010kr}) but
our main result survives small damping.
We leave open further extensive studies of the damping effects on the photon localization-delocalization.

We have explored the photon population dynamics in a system of two coupled circuit QED systems in the USC regime.
For $g\gtrsim\omega_0$, the recently observed localized photon
dynamics~\cite{Raftery:2014jk} gives way to quasi-equilibration dynamics. It
reveals a new qualitatively distinctive nonlinear behavior of the circuit QED
system.

\begin{acknowledgments}
MJH was supported by the ERC Synergy grant BioQ and the EU STREP DIADEMS and EQUAM and the IT R\&D program of MOTIE/KEIT [Grant No.~10043464(2012)].
The Python package QuTiP 2~\cite{Johansson13a} has been used for simulations.
MSK thanks the UK EPSRC. MSC acknowledges the support by the NRF of Korea (Grant No.~2015-003689).
\end{acknowledgments}

\bibliographystyle{apsrev4-1}
\bibliography{TwoCoupledCavities}

\end{document}